# Direct Measurement of Microwave Loss in Nb Films for Superconducting Qubits

B. Abdisatarov [1,2], D. Bafia [1], A. Murthy [1], G. Eremeev [1], H. E. Elsayed-Ali [2], J. Lee [1], A. Netepenko [1], C. P. A. Carlos [3], S. Leith [3], G. J. Rosaz [3], A. Romanenko [1], A. Grassellino [1]

[1] Fermi National Accelerator Laboratory, Batavia, Illinois 605010, USA

[2] Department of Electrical and Computer Engineering, Old Dominion University, Norfolk, Virginia 23529, USA and Applied Research Center, 12050 Jefferson Avenue, Newport News, Virginia 23606, USA

[3] CERN, European Organization for Nuclear Research, 1211 Geneva, Switzerland

**ABSTRACT**

Niobium films are a key component in modern two-dimensional superconducting qubits, yet their contribution to the total qubit decay rate is not fully understood. The presence of different layers of materials and interfaces makes it difficult to identify the dominant loss channels in present two-dimensional qubit designs. In this paper we present the first study which directly correlates measurements of RF losses in such films to material parameters by investigating a high-power impulse magnetron sputtered (HiPIMS) film atop a three-dimensional niobium superconducting radiofrequency (SRF) resonator. By using a 3D SRF structure, we are able to isolate the niobium film loss from other contributions. Our findings indicate that microwave dissipation in the HiPIMS-prepared niobium films, within the quantum regime, resembles that of record-high intrinsic quality factor of bulk niobium SRF cavities, with lifetimes extending into seconds. Microstructure and impurity level of the niobium film do not significantly affect the losses. These results set the scale of microwave losses in niobium films and show that niobium losses do not dominate the observed coherence times in present two-dimensional superconducting qubit designs, instead highlighting the dominant role of the dielectric oxide in limiting the performance. We can also set a bound for when niobium film losses will become a limitation for qubit lifetimes.

## INTRODUCTION

Superconducting qubits represent one of the leading technologies for quantum computing. The two-dimensional (2D) architecture typically includes superconducting resonators, feedlines, and Josephson junction-based structures patterned on a silicon or sapphire substrate [1, 2, 3]. In terms of materials, thin films of niobium (Nb) have been extensively employed as the superconducting base metal in these devices. [1, 2, 3].

Improving the coherence time of the 2D superconducting qubits is essential for enabling the practical implementation of quantum processors. Since the patterned structures include several materials, surfaces, and interfaces, all subjected to electromagnetic fields during operation, it is important to understand the contributions from various microwave loss channels [1, 3]. These include two-level system (TLS) -induced and conductive losses in the Nb film of the circuit. While it has been shown that bulk Nb metal exhibits very low microwave dissipation [4], the microwave losses in thin film Nb metal require further study. Previous studies of 2D structures focused on multifactor fitting and changing several circuit chip parameters [3, 5, 6, 7]. However, the intricate structure of superconducting quantum devices complicates the interpretation of computer simulations and requires assumptions used in these methods. A more accurate method to evaluate different contributions involves isolating different elements and measuring them independently. One of the approaches to achieve it is to study the Nb thin film material by itself as a three-dimensional (3D) resonator, as used by Romanenko's *et al.* [4]. They evaluated losses in bulk Nb 3D resonators and found that performance below 1.4 K and at low photon number is dominated by TLS present in the native oxide. Similarly, Checchin *et al.* [8] used a 3D cavity to isolate the silicon substrate, enabling them to directly measure the TLS loss participation ratio of both silicon and silicon oxide. This approach helps to better understand and quantify the loss contributions of different materials and interfaces of the qubit structure.

Furthermore, the structure and chemistry associated with a given thin film can be tuned significantly through annealing, which may result in reduced surface losses [4, 9, 10]. This reduction is crucial for achieving long coherence times and minimizing energy losses in superconductors. Annealing at different temperatures changes impurity levels, such as oxygen diffusion, surface oxide dissolution [9], and hydrogen degassing [11], thereby affecting the microstructure of the Nb film [10]. Microstructure changes include crystallite size, macrostrain, dislocation density, and other relevant parameters [10, 11, 12]. By clarifying the relationship between film structure, chemistry, and microwave loss, researchers can methodically design optimal treatments tailored to specific applications.

In this work, we isolate the Nb film from other qubit components and measure directly the conductive loss associated with the film and the TLS loss at the niobium-air interface. We perform this study by depositing a Nb film directly on the inner surface of a 3D Nb cavity. As such, we confine the microwave fields to the Nb film without concurrent contributions associated with components such as substrates or junctions. This Nb film has nanometer scale grains and residual resistivity ratio (RRR ~ 20) value comparable to Nb films deposited on substrates for qubits [1]. We find that the Nb film exhibits similar losses compared to bulk Nb. Moreover, we find that varying film material properties such as impurity concentration, macrostructure, RRR, crystallite size, and microstrain, does not impact microwave losses significantly. However, we find that the dissolution of the native oxide *via in-situ* vacuum annealing reduces the loss ten-fold. Our findings further highlight the deleterious impact of niobium oxides on qubit performance.

## EXPERIMENTAL SETUP

Nb films were grown on the inner surface of a 1.3 GHz elliptical single-cell, fine-grain (~50 μm), high RRR (~300) bulk Nb superconducting radiofrequency (SRF) cavity and on representative 10 × 10 × 3 mm Nb substrates. Film growth was achieved using a DC-biased HiPIMS film deposition system. Detailed information regarding the deposition conditions and parameters can be found in [13] and the supplementary materials. Using high resolution transmission electron microscopy (TEM), we confirmed a ~6.1 μm thick film with columnar grains ranging in size from 400 to 600 nm; some grains exceeded 1 μm. Additionally, the TEM examination indicates a measured thickness of approximately 3-4 nm for the $Nb_2O_5$ layer. This finding is consistent with previous studies on Nb film oxide layers [14] and aligns with the thickness of the pentoxide layer observed on bulk Nb [11] (see supplemental material).

Following the Nb film coating, the cavity underwent preparation for cryogenic testing. To mitigate the risk of stripping the Nb film from the substrate, ultrasonic cleaning was avoided, and instead, the cavity was rinsed with ultra-high purity water under a lower pressure of 50 bar. Subsequently, the cavity was assembled in an ISO4 clean room environment for cryogenic testing.

The characterization of the Nb films deposited on the substrates involved the utilization of TEM, time-of-flight secondary ion mass spectrometry (ToF-SIMS), and X-ray diffraction (XRD) techniques. Detailed methodologies are provided in the supplementary information.

The cryogenic testing initially utilized in Vertical Cavity Test Stand (VCTS), where the cavity was submerged in a bath of liquid helium and cooled to temperatures of 2.0 and 1.5 K. Measurements were conducted by employing a phase-lock loop to lock the cavity onto its resonance frequency, enabling the derivation of the intrinsic quality factor ($Q_0$) and field through power balance and decay analysis [15, 16, 17]. Following these tests, the cavity, still under vacuum, was equipped with six Cernox temperature sensors and integrated into a dilution refrigerator (DR) setup. Measurement schematics and details are found in the supplementary material. The $Q_0$ was measured by powering the cavity close to its $TM_{010}$ resonance frequency using a network analyzer, with measurements automated and conducted continuously over multiple days. During acquisition, the DR temperature was increased from 40 mK to 10 K.

To investigate the influence of microstructure of the film and impurities on losses in the quantum regime, we implemented various annealing procedures on both the coated cavity and the samples. Initially, an *in-situ* annealing was conducted at 340 °C for 1 hour, maintaining the vacuum inside the cavity. This was followed by subsequent vacuum furnace heat treatments at 600 °C and 800 °C for 3 hours. Following these latter heat treatments, the Nb film was exposed to air, thus reforming the native niobium oxide. Subsequently, the cavity underwent testing in the liquid helium dewar at 1.5 K while the annealed samples were subjected to comprehensive analysis using a range of material characterization tools.

## RESULTS

Figure 1 shows the results obtained from 40 mK to the 10 K in the DR at low electric fields ($E_{peak}$ < 3.8 V/m). The measured $Q_0$ of the Nb film cavity closely resembled that of a 3D bulk Nb cavity, as reported in [4]. This was surprising given the dramatic difference in RRR (~20 versus ~300) and grain boundary densities between the Nb film and the bulk Nb cavities. In fact, the Nb film cavity exhibited an even higher $Q_0$ below 1 K.

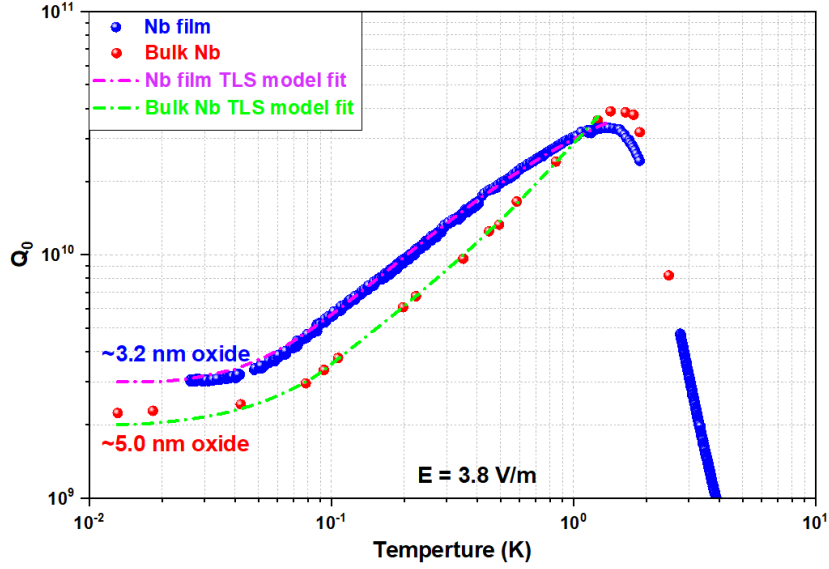

FIG 1. Temperature versus $Q_0$ of Nb film and bulk Nb cavities.

The $Q_0$ of the cavity was analyzed using the standard TLS model:

$$\frac{1}{Q(K)} = F\delta_{TLS}\, tanh\left(\alpha \frac{h\omega}{2kT}\right) + \frac{R_{res}}{G} \quad , \quad (1)$$

where $F$ is the filling factor obtained from simulation [4] ($3 \times 10^{-9}$ for 1.3 GHz tesla shaped cavity), $\delta_{TLS}$ is the loss tangent of the TLS at $T = 0$ K, $\alpha$ is the temperature measurement efficiency, $R_{res}$ is the residual resistance (1-3 n$\Omega$ for pure bulk Nb), $G$ is the geometry factor, $h$ is the Planck's constant, $k$ is the Boltzmann constant, and $T$ is the temperature in K. Best fits were achieved with $\delta_{TLS} = 0.12 \pm 0.01$, which is less than loss tangent of the bulk Nb (0.17) and $R_{res} = 4.25 \times 10^{-9}$ $\Omega$, which is slightly higher than the $R_{res}$ of the bulk Nb ($3.23 \times 10^{-9}$ $\Omega$ ). Weak links between grains, defects on the surface and trapped residual magnetic field are potential sources for this additional residual resistance [13].

After interrogating the as received cavity performance at low fields in the mK range, we studied how cavity performance evolves with material properties at higher fields and temperatures.

Figure 2 illustrates the relationship between the $Q_0$ and accelerating gradient ($E_{acc}$), for both bulk Nb and the as received Nb film cavities. It is evident that at high fields ($E_{acc} > 1$ MV/m) there is a significant degradation in the $Q_0$ observed for the Nb film cavity compared to the bulk Nb cavity. However, in the region where TLS loss dominates ($E_{acc} < 1$ MV/m) [18], the losses for both bulk Nb and Nb film cavities are comparable, consistent with our DR results.

During annealing between 100 and 400 °C, oxygen atoms diffuse within the Nb material, following the dissolution of oxide layer near the surface [9, 11]. *In-situ* annealing of the Nb film cavity for 1 hour at 340 °C results in complete dissolution of pentoxide (see supplementary material). After this treatment, the 3D resonator was tested without re-exposing it to ambient conditions. The field dependent TLS loss was almost eliminated, as demonstrated in Figure 2, with the $Q_0$ saturation level subsequently increasing approximately ten-fold.

The $Q_0$ as a function of the $E_{acc}$ after vacuum furnace annealing is shown in Figure 2. Annealing caused changes in impurity levels and microstructure of the film, leading to significant changes in the $Q_0$ and field dependency of the Nb film cavity for $E_{acc} > 1$ MV/m. However, these changes did not notably affect the losses in the TLS loss dominated region.

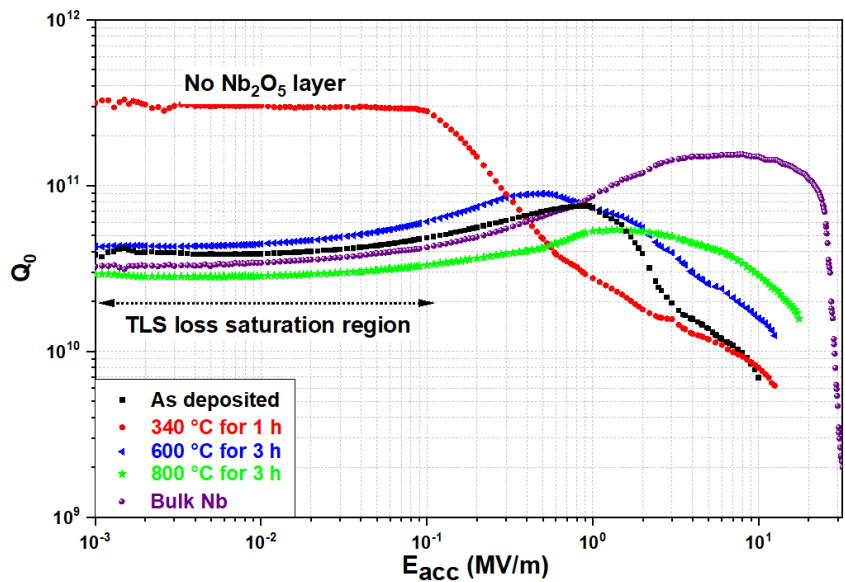

FIG 2. $Q_0$ versus $E_{acc}$ characteristics of bulk Nb and Nb film cavities before and after annealing procedures.

High temperature vacuum furnace annealing at 600 °C and 800 °C for 3 hours after *in-situ* annealing modified both the microstructure and impurity contents of the Nb film cavity. The ToF-SIMS depth profiling results (shown in Figure 3) highlight the surface characteristics of the film, particularly within a few nanometers. Sequential *in-situ* vacuum furnace annealing at 340 °C for 1 hour followed by annealing at 600 °C then 800 °C for 3 hours led to a reduction in oxygen concentration and complete pentoxide dissolution. Notably, upon exposure to air, the oxide layer reformed on the film, albeit not to its original thickness even after a week-long exposure to air in the laboratory environment, as depicted in Figure 3a.

Expulsion of hydrogen atoms from Nb was observed following each annealing at 600 °C and 800 °C, with no subsequent increase observed after a week of air exposure (Figure 3b), indicating hindered diffusion due to the regenerated oxide layer. Carbon levels remained relatively constant at the film surface, whereas internal carbon levels increased post-annealing (Figure 3c). Furthermore, nitrogen levels exhibited an increase after annealing at 340 °C, followed by a small decrease with higher temperature annealing (Figure 3d).

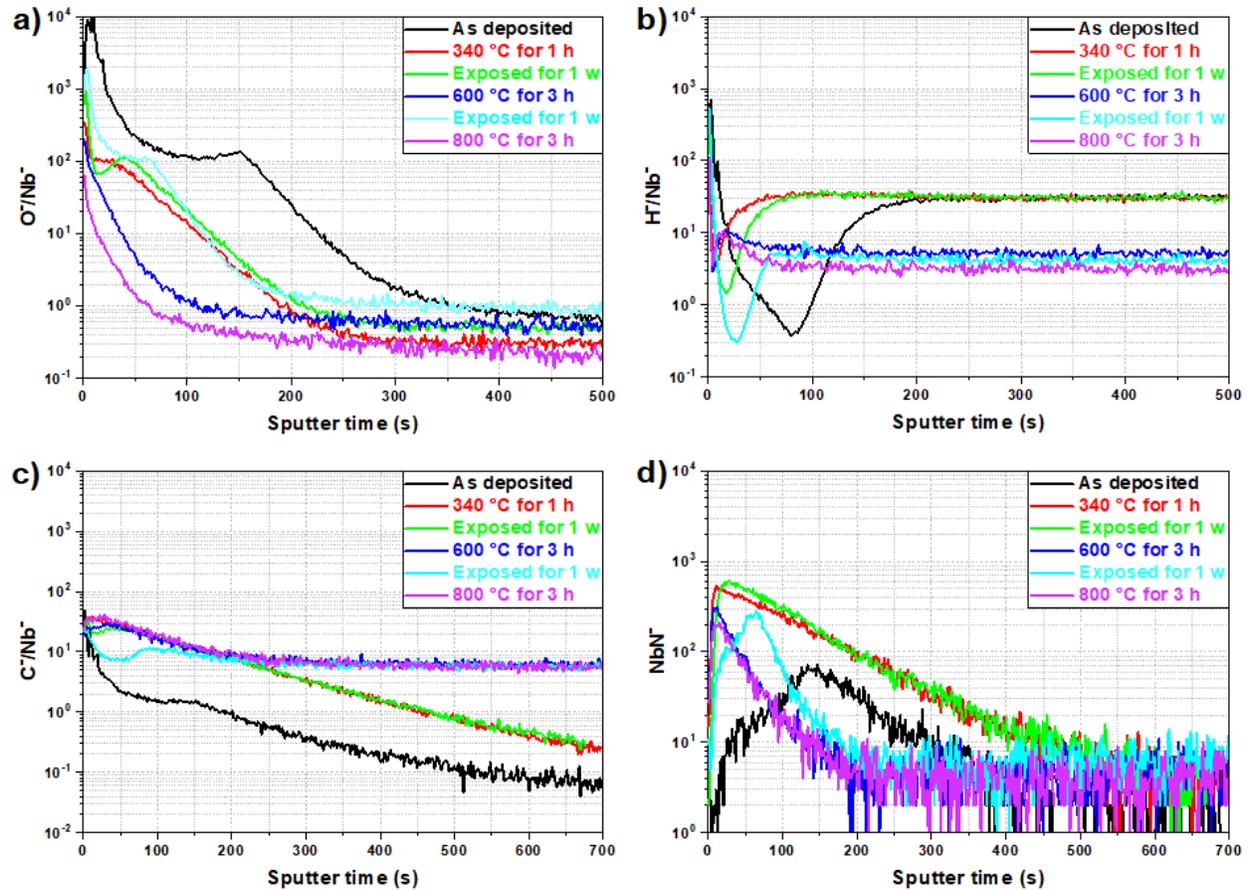

FIG 3. ToF-SIMS depth profiling before and after annealing procedures. Oxygen diffusion (a). Hydrogen degassing (b). Carbon level change (c). Niobium nitride (d).

XRD analysis revealed notable alterations in the microstructure of the Nb film after subsequent annealing. The Nb film samples display clear polycrystalline texturing, predominantly with the Nb (110) orientation, with the detection of Nb (200), Nb (211), and Nb (220) oriented structures. The average crystallite size and microstrain are calculated by using modified Williamson-Hall method [19] and summarized in the Table 1. After each annealing process, there was a noticeable increase in crystallite size, coupled with a decrease in microstrain. More details are found in the supplementary material.

Table 1: Microstructure change of Nb film after annealing procedures.

|  | Crystallite size (nm) | Microstrain ($\varepsilon$) |
| --- | --- | --- |
| As deposited | 32.18 | 3.41 |
| 340 °C for 1 h | 33.95 | 3.23 |
| 600 °C for 3 h | 35.35 | 3.10 |
| 800 °C for 3 h | 41.22 | 2.66 |

We extracted the mean free path within the RF layer after each treatment using Slater's theorem and the Halbritter code [20]. We find the mean free path of the as received film to be 61 nm, corresponding to RRR ≈ 20, which is comparable to RRR values expected for the best Nb films deposited by HiPIMS technique [13]. Following all annealing steps, the mean free path of the Nb film was calculated to be approximately 134 nm, which corresponds to RRR ≈ 50.

## DISCUSSION

The investigation into the loss mechanisms in the Nb film at the quantum regime involved manipulating microstructure, surface oxide layer and impurity levels of the film. It was found that the small grains and low RRR characteristic of Nb films do not significantly impact losses at the low fields and temperatures relevant for the quantum regime. This observation is supported by comparable $Q_0$ observed in both low temperature DR measurements and high field VCTS measurements between fine grain bulk Nb (50 μm) and small grain Nb film (400-600 nm) [13]. Moreover, the TLS loss model fits yielded similar values, with a slightly higher $Q_0$ observed in the Nb film cavity, potentially attributable to a slightly thinner oxide layer compared to that observed for a bulk Nb cavity. Specifically, the loss tangents were found to be almost identical, with measurements yielding values of approximately 0.17±0.02 for bulk Nb and 0.12±0.01 for Nb film at weak fields. Additionally, TLS loss saturation levels were nearly identical for large grain bulk Nb cavity and small grain Nb film cavity before and after annealing procedures, except from *in-situ* annealing. This suggests that despite significant changes in crystallite size, microstrain, and dislocation density due to annealing procedures, losses in the TLS dominated region remained consistent. Notably, the annealing procedures decreased surface resistance in the non-TLS dominated region ($E_{acc}$ > 1 MV/m) but did not affect it at TLS-dominated fields.

The oxide layer emerged as a significant factor in driving TLS loss in the Nb films. Pentoxide dissolution after *in-situ* annealing at 340 °C for 1 hour resulted in no observed TLS loss (Figure 2). However, regrowth of the oxide layer post-annealing introduced variability, highlighting the importance of understanding oxide layer dynamics, in line with observations made in Bafia *et al.* [9]. The thickness of the oxide layer, typically ranging from 3 to 5 nm on the bulk Nb surface [11, 14], was measured as 3.2 nm for the Nb film. This thickness variation may explain the slightly higher TLS loss in bulk Nb, where the oxide layer could be thicker than 3.2 nm.

A substantial decrease in hydrogen levels was observed after high-temperature annealing. However, this did not correlate with changes in the loss tangent or the saturation level of $Q_0$ in the quantum regime. Even with the lowest hydrogen level (Figure 3b) recorded after the highest temperature annealing, no significant effects on TLS loss were detected (Figure 2). Similarly, variations in NbN content, which peaked after *in-situ* annealing at 340 °C for 1 hour (Figure 3d), did not impact TLS loss, despite differences observed post-annealing. The highest NbN content showed no field-dependent TLS loss, and $Q_0$ was saturated at $3 \times 10^{11}$ (Figure 2). Higher temperature annealing significantly alter the NbN level. However, the saturation level of the $Q_0$ changed after annealing at 600 °C and 800 °C (Figure 2), suggesting a nuanced relationship between NbN content and TLS loss. Changes in carbon levels also showed no significant effect on TLS loss (Figure 2), with conflicting results after annealing; despite increased carbon levels (Figure 3c), TLS losses remained low after *in-situ* annealing at 340 °C for 1 hour. After annealing at 600 °C, the loss tangent was smaller and the saturation level higher than for the as deposited cavity, contradicting expectations based solely on carbon measurements.

While annealing procedures altered impurity levels, the mean free path and RRR increased significantly. However, the TLS loss increased, and $Q_0$ saturation level decreased after annealing at 800 °C for 3 hours in a vacuum furnace, which increased RRR, compared to the as-deposited condition. This underscores that lower RRR of sputtered films does not result in higher microwave losses in the quantum regime. An important finding of our study is that the regrowth process and thickness of the oxide layer significantly constrain coherence time in superconducting quantum devices. Specifically, we observed that variations in the surface oxide layer of niobium films primarily dictate the extent of TLS losses.

Lastly, the contribution of different oxide phases, such as $NbO_2$ and $NbO$, to TLS loss warrants further analysis, especially considering field dependent loss measurements conducted at 1.5 K, which is slightly above the superconducting critical temperature of $NbO$ (1.38 K). The role of these phases could provide deeper insights into the observed temperature dependent TLS losses.

## CONCLUSION

In our investigation of the microwave loss characteristics of Nb films within the quantum regime, several pivotal findings have emerged. Despite the small grain size and relatively low RRR of HiPIMS-prepared Nb films, our observations indicate a remarkable similarity in microwave dissipation to that of large grain and high RRR bulk Nb SRF. Notably, the grain size and RRR appears to have minimal impact on losses, as evidenced by the comparable $Q_0$ observed for large-grain, high RRR bulk Nb and small-grain, low RRR Nb films. Moreover, variations in crystallite size, macrostrain, and dislocation densities exhibit negligible effects on microwave losses at low field and temperatures.

Furthermore, our investigation into hydrogen, nitrogen, and carbon levels revealed no significant influence on TLS loss, underscoring the robust nature of Nb films. However, we highlight the key role played by the oxide layer in TLS loss and saturation levels, with dynamics associated with pentoxide dissolution and regrowth notably impacting TLS loss characteristics. Further studies of the oxide phases, such as $NbO_2$ and $NbO$, are needed to improve our understanding of the effect of surface oxides on Nb thin film SRF cavity operation in the quantum regime. Present findings contribute to the understanding of microwave dissipation in Nb films and offer avenues for future research aimed at enhancing performance and reliability for quantum applications.

## ACKNOWLEDGMENT

This work was supported by the U.S. Department of Energy, Office of Science, National Quantum Information Science Research Centers, Superconducting Quantum Materials and Systems Center (SQMS) under contract No. DE-AC02-07CH11359.## REFERENCES

[1] M. Kjaergaard, M. E. Schwartz, J. Braumuller, P. Krantz, J. I.-J. Wang, S. Gustavsson, and W. D. Oliver, "Superconducting qubits: Current state of play", Annual Review Condensed Matter Physics, 2020.
[2] W. D. Oliver, "Quantum Information Processing", IFF Spring School, 2013.
[3] C. Wang, C. Axline, Y. Y. Gao, T. Brecht, Y. Chu, L. Frunzio, M. H. Devoret, and R. J. Schoelkopf, "Surface participation and dielectric loss in superconducting qubits", Applied Physics Ltters, 2012.
[4] A. Romanenko, R. Pilipenko, S. Zorzetti, D. Frolov, M. Awida, S. Belomestnykh, S. Posen, and A. Grassellino, "Three-dimensional superconducting resonators at T < 20 mK with photon lifetimes up to τ = 2 s", Physical Review Applied, 2020.
[5] J. M. Gambetta, C. E. Murray, Y.-K.-K. Fung, D.T. McClure, O. Dial, W. Shanks, J. W. Sleight, and M. Steffen, "Investigating Surface Loss Effects in Superconducting Transmon Qubits", IEEE Transaction on Applied Superconductivity, 2017.

## SUPPLEMENTARY INFORMATION

### A. Film deposition

Nb films were grown using a DC-biased HiPIMS film deposition system in balance configuration with a high purity, residual resistivity ratio (RRR) > 300, Nb target. Before deposition, the coating apparatus and the cavity were baked out at 200 °C for 48 hours. The base pressure of the whole setup reached about $5 \times 10^{-10}$ mbar before deposition. The sample were coated at $3 \times 10^{-3}$ mbar ultra-pure krypton pressure, the substrate temperature of 150 °C, a bias voltage of -75 V, and the main pulse length of 200 μs. The deposition was done for 6 hours at the peak power of 1.2 kW, which resulted in about 6 μm Nb film. To evaluate the microstructure and composition, Nb samples, made out of high purity Nb sheet used for cavity fabrication, were coated under the conditions described above. The deposition was done in the same HiPIMS system used for cavity deposition, using specially made sample holder cavity to which the samples were mounted [11]. Nb samples, used as the substrates for film deposition, were cut to $10 \times 10$ mm size from 3 mm thick fine grain Nb sheet with the RRR of about 300. The samples were electropolished for 100 μm in EP solution, vacuum annealed at 800 °C for 3 hours, then electropolished for 25 μm in EP solution. After chemical polishing samples were first ultrasonically cleaned for 30 minutes at 50 °C in Liquinox solution and then for another 30 minutes at 50 °C in ultra-pure water. They were dried overnight in the cleanroom and bagged. Prior to coating, the samples were rinsed with demineralized water and ultrapure alcohol, followed by dry air blowing. The samples were then assembled in the cavity and the coating setup was assembled in a cleanroom.

### B. Film characterization

The microstructure of the Nb film deposited on the Nb substrate was examined using transmission electron microscopy (TEM). TEM characterization was conducted using a JEOL aberration-corrected Grand ARM-200 microscope. TEM specimens were prepared using a Thermo Scientific Helios 5 DualBeam SEM (Scanning Electron Microscope), following a standard protocol for TEM specimen preparation. Measurements were performed on three different samples at randomly selected sites using a 5 kV accelerating voltage. Moreover, TEM images were used to measure the thickness of the oxide layer on the surface of the Nb film. The High-resolution TEM images were analyzed and processed using the Gatan Micrograph Suite version 3.0.

The high-resolution TEM analysis reveals columnar growth grains within the film, with an average grain size ranging from 400 to 600 nm and occasional grains exceeding 1 µm, as shown in Figure 4a. The film thickness was measured to be approximately 6.1 µm. Notably, nanometer-sized grains are observed at the interface between bulk Nb substrate and the Nb film, with evident grain coarsening in the direction towards the surface. Additionally, the high-resolution TEM examination indicates a measured thickness of approximately 3-4 nm for the $Nb_2O_5$ layer. The morphology of the pentoxide layer closely mirrors the surface characteristics of the Nb film, as illustrated in Figure 4b.

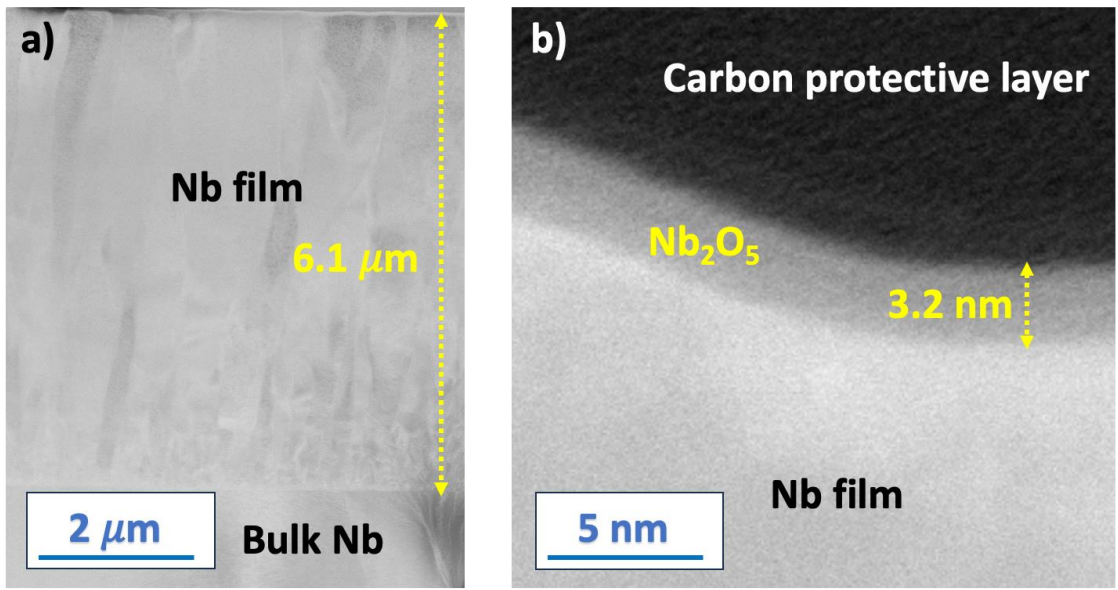

FIG 4. High resolution TEM image of the Nb film. Thickness of the film was 6.1 µm (a) while thickness of the $Nb_2O_5$ was 3.2 nm (b).

Chemical constituents present within the Nb films were analyzed using IONTOF dual beam time-of-flight secondary ion mass spectrometry (ToF-SIMS). ToF-SIMS measurements were conducted with an area of analysis of 200 × 200 μm inside a sputtering area of 500 × 500 μm. A liquid bismuth ion beam ($Bi^+$) with an energy of 30 keV was used for sputtering, while a cesium ion gun ($Cs^+$) with an energy of 500 eV was employed for ionization during depth profiling. Analysis was performed at a pressure of approximately $2 \times 10^{-10}$ mbar.

Before we analyzed the depth profile obtained from SIMS measurement, we used special feature of the software used to analyze the data. A crucial feature of this TOF-SIMS systems is the ability to utilize full 3D imaging information to ensure that the obtained depth profiles are not affected by any artifacts, such as particles of carbon and oxides. on the surface. The software can exclude the particles on the surface before the integration. In one of the measurements an unexpected artifacts and particles appeared like a bright spotlight. Our SIMS analysis showed that they are carbon particles. These appeared as three bright areas which were removed to obtain more homogeneous surface as shown in Figure 5a and 5b.

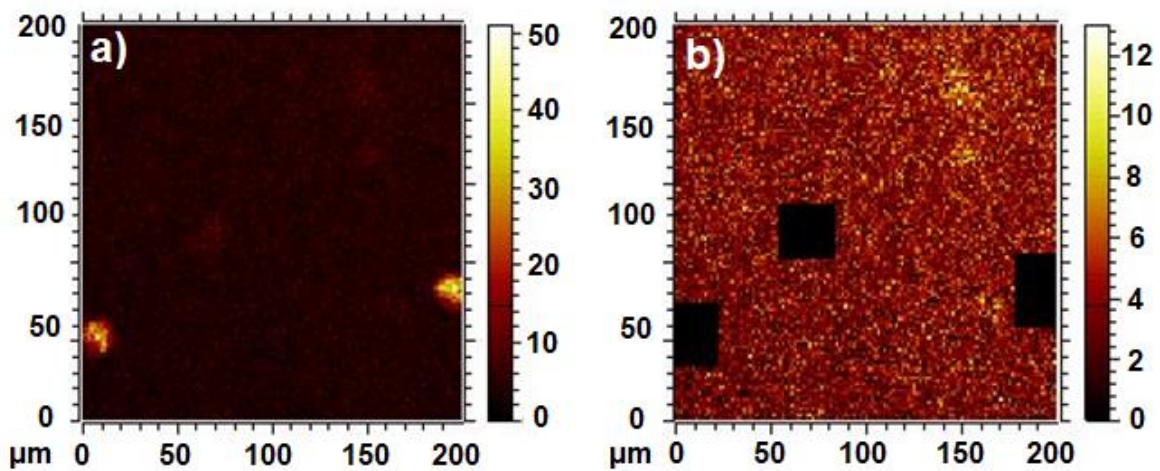

FIG 5. ToF SIMS 3D image. Unexpected artifacts appeared as bright spots on the surface (a).

Unexpected artifacts excluded to analyze the data more precisely (b).

The ToF SIMS depth profile of the Nb film before and after in-situ annealing for 1 hour at 340 °C is illustrated in Figure 6. Before annealing $Nb_2O_5$ layer was sputtered in 200 seconds which corresponds to 3-4 nm. After annealing pentoxide layer almost eliminated.

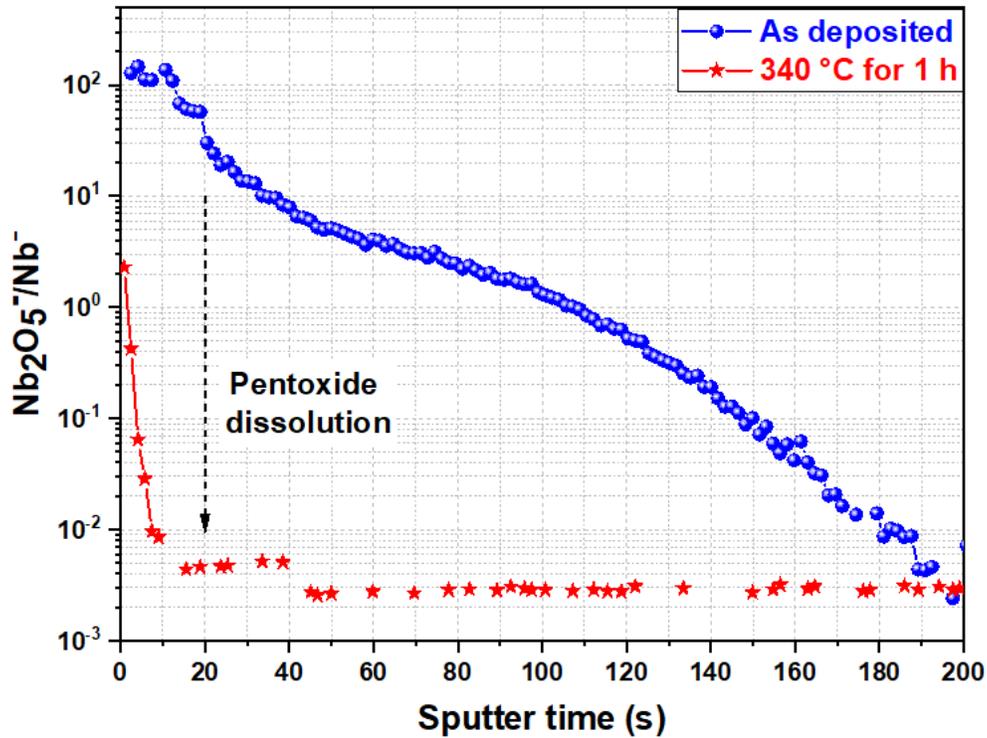

FIG 6. ToF-SIMS measurement of Nb film sample before and after *in-situ* annealing for 1 hour at 340 °C. Following annealing, elimination of the pentoxide layer is observed.

The crystallographic orientation, crystallite size, lattice strain, and dislocation density of the samples were analyzed using the Rigaku Smartlab Gen2 X-ray diffraction instrument, with Cu Kα radiation and λ = 1.54 Å. To examine the surface layer of the Nb film, an incoming X-ray grazing incident angle of 1 ° is selected. The detector scans between 30 ° and 90 ° with a step size of 0.02 ° and a scan rate of 0.5 ° per minute. XRD results showed that the film is a polycrystalline texturing, with the Nb (110), Nb (200), Nb (211), and Nb (220) oriented structures as shown in Figure 7.

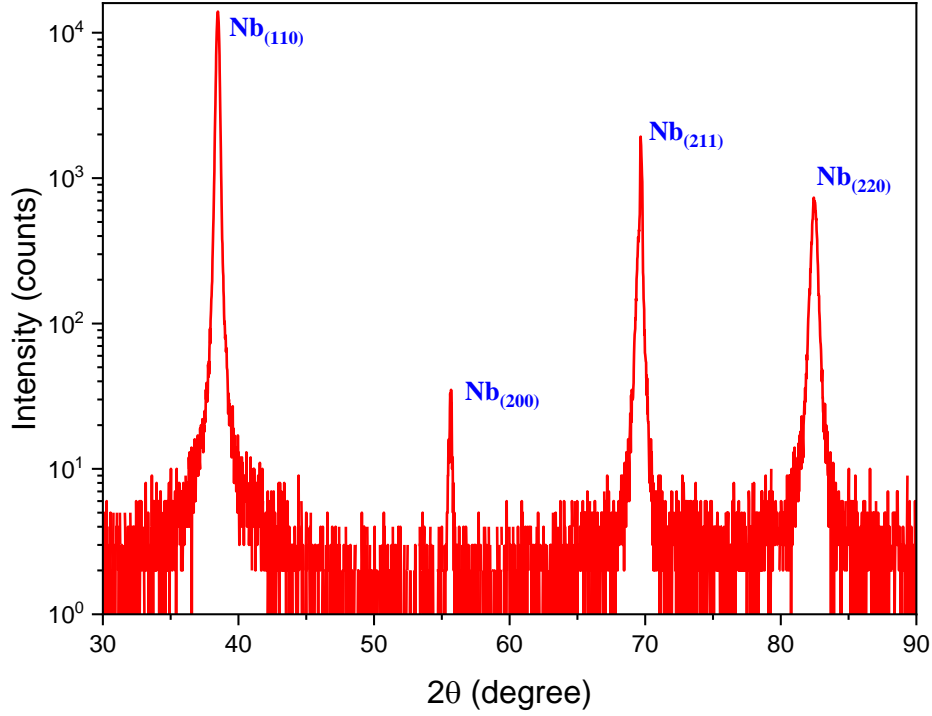

FIG 7. XRD pattern of Nb thin film shows Nb (110), Nb (200), Nb (211), and Nb (220) peaks.

### C. DR measurement setup

Low field performance of the cavity was tested in DR. The electrical field within the cavity was determined using the equation:

$$E = k\sqrt{P_W Q_{ext2}} \qquad (2)$$

where $E$ represents the electrical field in V/m, $k$ value is the specific value for SRF cavity which is 88.5 $\Omega^{1/2}$/m for 1.3 GHz single cell elliptical Nb cavity, $P_W$ is the transmitted power in Watts and $Q_{ext2}$ is the external quality factor of the transmitted coupler which is $1.4 \times 10^{-12}$ for our cavity test. During testing, the maximum power in the cavity reached $1.3 \times 10^{-15}$ W, corresponding to $E$ = 3.8 V/m.

The cavity integrated into the DR setup, as shown in Figure 8. Subsequently, the quality factor of the cavity was assessed via power decay measurements at various temperature points ranging from 40 mK to 10 K. Temperature monitoring was achieved using six Cernox sensors. $Q_0$ determination was performed by powering the cavity close to its $TM_{010}$ resonance frequency using a network analyzer, with measurements automated and conducted continuously over multiple days. The MATLAB code was developed to determine the best fit for each power decay profile to extract the decay time constant and corresponding $Q_0$. Furthermore, the cavity's field was estimated based on the transmitted power measurements, adjusted for microwave component losses between the cavity and the power meter. Temperature assessment involved averaging the temperatures measured at two flange points.

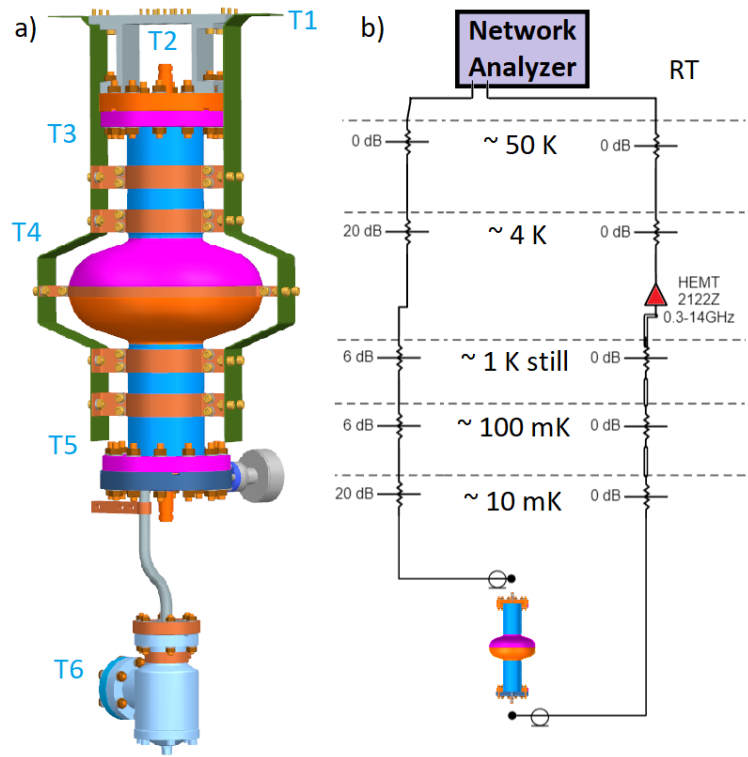

FIG 8. a) Shows positions of the temperature sensors on the cavity: T1 was placed on the mixing chamber plate; T2 was attached to the input power N-connector; T3 (was connected to the top flange of the cavity; T4 was attached to the cooling frame at the equator; T5 was attached to the bottom flange of the cavity; T6 was attached to the right-angle valve. b) Microwave setup for the measurements.